\documentclass[runningheads]{llncs}

\usepackage{hyperref}

\usepackage{scalerel}
\usepackage{tikz}
\usetikzlibrary{svg.path}
\definecolor{orcidlogocol}{HTML}{A6CE39}
\tikzset{
  orcidlogo/.pic={
    \fill[orcidlogocol] svg{M256,128c0,70.7-57.3,128-128,128C57.3,256,0,198.7,0,128C0,57.3,57.3,0,128,0C198.7,0,256,57.3,256,128z};
    \fill[white] svg{M86.3,186.2H70.9V79.1h15.4v48.4V186.2z}
    svg{M108.9,79.1h41.6c39.6,0,57,28.3,57,53.6c0,27.5-21.5,53.6-56.8,53.6h-41.8V79.1z M124.3,172.4h24.5c34.9,0,42.9-26.5,42.9-39.7c0-21.5-13.7-39.7-43.7-39.7h-23.7V172.4z}
    svg{M88.7,56.8c0,5.5-4.5,10.1-10.1,10.1c-5.6,0-10.1-4.6-10.1-10.1c0-5.6,4.5-10.1,10.1-10.1C84.2,46.7,88.7,51.3,88.7,56.8z};
  }
}
\newcommand\orcidicon[1]{\href{https://orcid.org/#1}{\mbox{\scalerel*{\begin{tikzpicture}[yscale=-1,transform shape]\pic{orcidlogo};\end{tikzpicture}}{|}}}}

\usepackage{graphicx}
\usepackage{multirow}
\usepackage[numbers,sort&compress]{natbib}
\usepackage{float}
\usepackage{stackengine}
\newcommand\xrowht[2][0]{\addstackgap[.5\dimexpr#2\relax]{\vphantom{#1}}}
\begin{document}
\title{ABCD Neurocognitive Prediction Challenge 2019: Predicting individual residual fluid intelligence scores from cortical grey matter morphology}
\author{Neil P. Oxtoby\inst{1,*}\protect\orcidicon{0000-0003-0203-3909} \and
Fabio S. Ferreira\inst{1,2,*}\protect\orcidicon{0000-0002-0977-2539} \and
Agoston Mihalik\inst{1,2}\protect\orcidicon{0000-0002-4510-4933} \and
Tong Wu\inst{1,2}\protect\orcidicon{0000-0002-7468-2249} \and
Mikael Brudfors\inst{1,3}\protect\orcidicon{0000-0002-2884-2336}\and 
Hongxiang Lin\inst{1}\protect\orcidicon{0000-0001-6643-327X} \and
Anita Rau\inst{1,3}\protect\orcidicon{0000-0002-4759-2846} \and
Stefano B. Blumberg\inst{1}\protect\orcidicon{0000-0002-7150-9918} \and
Maria Robu\inst{1,3}\protect\orcidicon{0000-0003-0106-0542} \and
Cemre Zor\inst{1,2}\protect\orcidicon{0000-0002-6141-2610} \and
Maira Tariq\inst{1}\protect\orcidicon{0000-0002-2826-4046} \and
Maria Del Mar Estarellas Garcia\inst{1} \and
Baris Kanber\inst{5}\protect\orcidicon{0000-0003-2443-8800} \and
Daniil I. Nikitichev\inst{1,3}\protect\orcidicon{0000-0001-5877-9174} \and
Janaina Mourao-Miranda\inst{1,2}\protect\orcidicon{0000-0002-3309-8441}}

\authorrunning{N.P. Oxtoby \& F.S. Ferreira, \textit{et al.}}
\titlerunning{Predicting residual fluid intelligence from cortical grey matter morphology}
\institute{%
Centre for Medical Image Computing (CMIC), \newline{}Department of Computer Science \&\newline{}Department of Medical Physics and Biomedical Engineering, \and
Max Planck UCL Centre for Computational Psychiatry and Ageing Research, \and
The Wellcome Centre for Human Neuroimaging \and
Wellcome/EPSRC Centre for Interventional and Surgical Sciences (WEISS), \and
Department of Clinical and Experimental Epilepsy,\newline{}Queen Square Institute of Neurology; \\ \vspace{3mm}
University College London, Gower Street, London, WC1E 6BT, United Kingdom \\ \vspace{3mm}
\textasteriskcentered\ These authors contributed equally to this work.}

\maketitle

\begin{abstract}
We predicted residual fluid intelligence scores from T1-weighted MRI data available as part of the ABCD NP Challenge 2019, using morphological similarity of grey-matter regions across the cortex. Individual structural covariance networks (SCN) were abstracted into graph-theory metrics averaged over nodes across the brain and in data-driven communities/modules. Metrics included degree, path length, clustering coefficient, centrality, rich club coefficient, and small-worldness. These features derived from the training set were used to build various regression models for predicting residual fluid intelligence scores, with performance evaluated both using cross-validation within the training set and using the held-out validation set. Our predictions on the test set were generated with a support vector regression model trained on the training set. We found minimal improvement over predicting a zero residual fluid intelligence score across the sample population, implying that structural covariance networks calculated from T1-weighted MR imaging data provide little information about residual fluid intelligence.
\end{abstract}

\section{Introduction}\label{INTRO}
Establishing the neurobiological mechanisms underlying intelligence is a key area of research in Neuroscience \cite{Goriounova2019}. A strong correlation has been observed between cognitive ability measured at a very young age with the socioeconomic status \cite{Foverskov2017}, as well as longevity and health \cite{Lam2017}, at an older age. Moreover, intelligence has been shown to be very stable from young to old age in the same individuals \cite{Deary2007}\cite{Deary2013}. Thus understanding the mechanisms of cognitive abilities has implications for health of the general population and can be used to enhance such abilities, for example through education or environment \cite{Gottfredson1997}.  

Neuroimaging plays a key role in advancing our knowledge of the neurological mechanisms of intelligence. Several brain-imaging studies have shown the link between brain features and intelligence, including a positive correlation with cortical volume and thickness, specifically in the frontal and temporal regions \cite{HulshoffPol2006,Narr2007,Choi2008,Karama2011,Jung2007}. A link has also been observed between intelligence and the structural integrity of white matter \cite{Penke2012} and the function integrity of the temporal, frontal and parietal cortices \cite{Wang2009}. Studies have also involved both adult and children \cite{Muetzel2015,Yu2008}.  The ABCD NP Challenge asks the question ``How predictable is fluid intelligence from brain imaging data?'' To answer this, we took a data-driven, exploratory approach of trying many models and image-based features --- starting with a hackathon led by the UCL Centre for Medical Image Computing (CMIC). CMIC aims to make an impact on key medical challenges facing 21st century society through performing world-leading research on problems in medical imaging and image-analysis. Our expertise extends from feature extraction/generation through to image-based modelling \cite{Oxtoby2017,Young2014}, machine learning \cite{Schrouff2018,Blumberg2018}, and beyond. The hackathon took place one afternoon in February 2019 and involved researchers across research groups in UCL CMIC, in addition to colleagues from the affiliated UCL Wellcome Centre for Human Neuroimaging, UCL Department of Clinical and Experimental Epilepsy, and Max Planck UCL Centre for Computational Psychiatry and Ageing Research. Regular followup progress meetings followed the hackathon.

The brain is a complex organ widely touted as operating as a cliquish small-world network \cite{Bassett2006}, although this may not be the whole story \cite{Bassett2017}. The ABCD NP Challenge lacks the diffusion MRI data necessary to estimate anatomical connectivity via tractography. However, it is possible to quantify morphological similarity of an individual's cortex using a graph called a ``structural covariance network'' (SCN), which can be used to distinguish between clinical groups \cite{Tijms2013}. We calculate SCNs for each individual in the ABCD NP Challenge data set and input them as features to train predictive models of residual fluid intelligence (rFIQ). 

The paper is structured as follows. The next section describes the challenge data and our methods. Section \ref{RESULTS} presents our results which we discuss in section \ref{DISCUSSION} then conclude.

\section{Methods}\label{METHODS}
\subsection{Data}
The ABCD NP Challenge data consists of a cross-section of imaging data and intelligence scores for children aged 9--10 years. The T1-weighted MRI data was acquired using the protocol detailed on the challenge website \cite{ABCDWEB} and in \cite{Casey2018a}, and split into training ($N=3739$), validation ($N=415$), and test ($N=4515$) sets. The training and validation sets also include scores of fluid intelligence, which the ABCD Study measures using the NIH Toolbox Neurocognition battery \cite{Akshoomoff2013}. For the challenge, fluid intelligence was residualized to remove dependence upon brain volume, data collection site, age at baseline, sex at birth, race/ethnicity, highest parental education, parental income, and parental marital status. While we understand the motivation --- the challenge is to predict intelligence from imaging --- this pre-residualization choice in the challenge design is somewhat limiting because it completely removes any ability to include covariance of these factors with image-based features. The MRI data provided was already in pre-processed form. Pre-processing included skull-stripping, removing noise, correcting for field inhomogeneities \cite{Hagler2018,Pfefferbaum2018} and affine alignment of all images to the SRI24 adult brain atlas \cite{Rohlfing2010}. The SRI24 segmentations and corresponding volumes were also provided. Unsurprisingly, the regional volumes were not predictive of a target that had been adjusted for total brain volume.

\subsection{Structural Covariance Network Features}
\begin{figure}[H]
\begin{center}
\includegraphics[width=0.9\columnwidth]{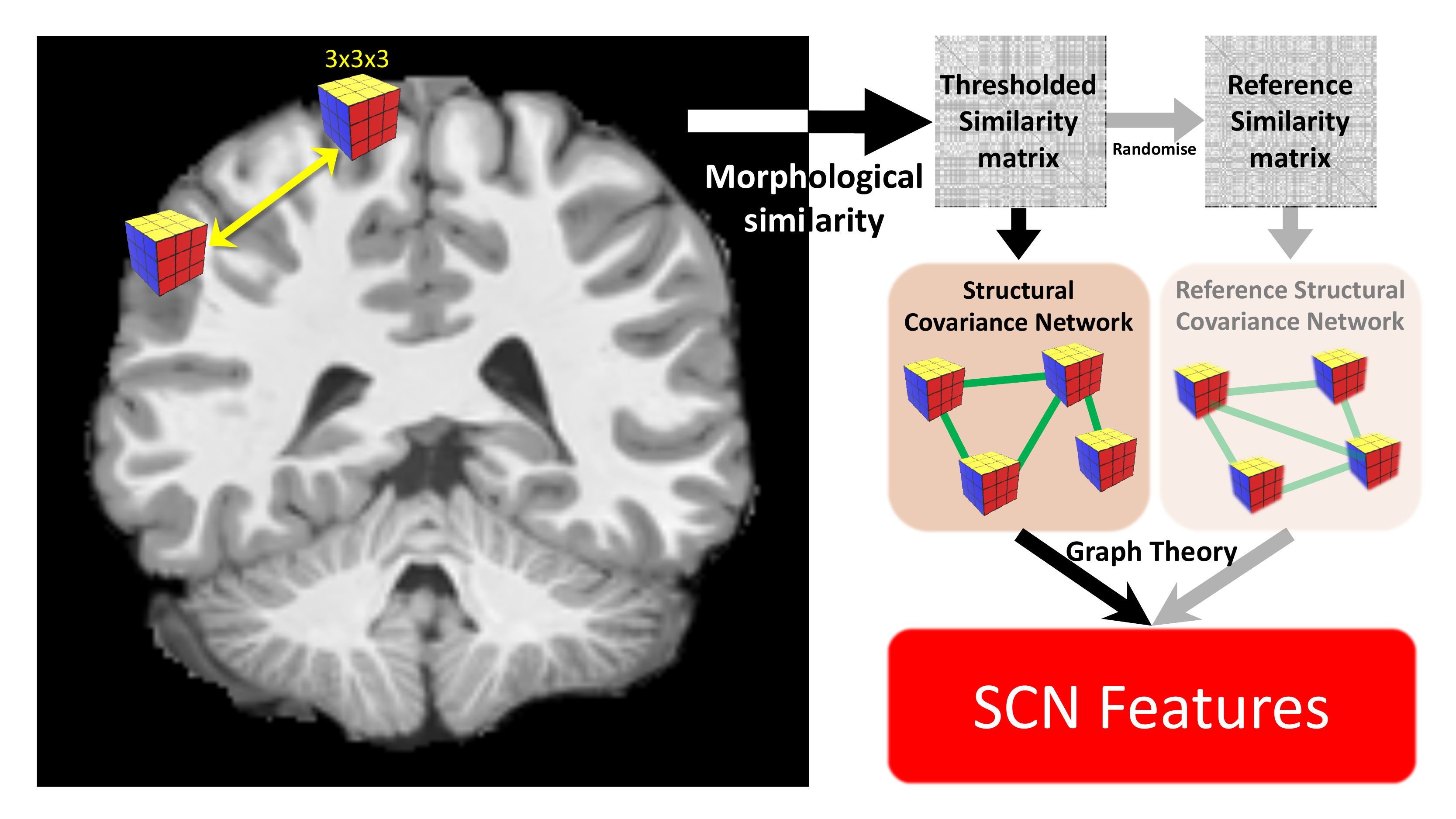}
\caption{Structural Covariance Network feature-generation pipeline.}\label{Fig-0}
\end{center}
\end{figure}

It has been shown that cortical morphology is predictive of cognitive deficits in individuals with Alzheimer's disease \cite{Tijms2013}. We wanted to explore whether the same could be said for predicting intelligence, so we generated a structural covariance network (SCN) following \cite{Lawrie2012} (code available on GitHub) for each individual in the ABCD NP Challenge data set. The SCN is a graph where the nodes are small cortical regions (3 voxels cubed) and the edges quantify structural similarity (morphology) between nodes. From each SCN we generated nodal graph-theory features using the Brain Connectivity Toolbox \cite{Rubinov2010}, which were then averaged across the brain and also within each of the largest three modules (communities) of the graph. We also considered measures of variation in these features (standard deviation and median-absolute deviation). Our 26 features include small-worldness, rich club coefficient, path length, node degree, clustering coefficient, and betweenness centrality (Table~\ref{Table-1}). See Figure~\ref{Fig-0} for a graphical representation of the pipeline.

Generating approximately ten thousand SCNs and corresponding graph-theory features is an intensive computational task. When the pipeline failed for a given individual, or time was not permitting (such as the late addition of 868 additional test subjects), this resulted in missing data. For these few individuals ($\leq8\%$: Table~\ref{Table-1}) we inserted a prediction of zero (nominally the mean). 

\subsection{Predictive Models}\label{predmod}
We trained two models to predict rFIQ from features based on morphological similarity. The first was the event-based model (EBM) of progression \cite{Young2014,Fonteijn2012}. The second was support-vector regression (SVR) \cite{Drucker:1996:SVR:2998981.2999003}. We trained each model on data from the training set, and assessed performance using MSE on the validation set (Table~\ref{Table-1}). The best-performing model (SVR) was used to generate our submission to the challenge: predictions for the test set. 

The EBM learns a discrete sequence of progression events from normal/low state to abnormal/high. It was designed for neurodegenerative diseases but can be applied to any monotonic phenomenon. Here we define low rFIQ as more than one standard deviation (std) below the mean and high rFIQ as more than one std above the mean. If rFIQ is a monotonic function of structural covariance, then the EBM should be able to find a probabilistic sequence of events that represent this function. ``Events'' are structural covariance graph-theory features, and they must differ statistically between low-rFIQ and high-rFIQ for them to be included in the model --- otherwise they contain no ``signal'' for this trajectory. We excluded features that “did not pass” ($p>0.10$) the Mann-Whitney U test of the null hypothesis that the distributions (low/high rFIQ) are equal. EBM stage and rFIQ score was input into a Kernel Ridge Regression model (default parameters, scikit-learn: \cite{Pedregosa2011}) to make the predictions. 

The SVR was run in PRoNTo version 3 (Pattern Recognition for Neuroimaging Toolbox) \cite{PRONTO,Schrouff2013} --- a software toolbox of pattern recognition techniques for the analysis of neuroimaging data. Model performance on the training set was assessed using 5-fold nested cross-validation (i.e. the internal and external loops had 5 folds) to optimise the penalty parameter C (we use 6 different logarithmically-spaced values: 0.01, 0.1, 1, 10, 100 and 1000) and compute the MSE per fold, which were averaged across folds to compute the final prediction error (Table~\ref{Table-2}).

\begin{table}[H]
\caption{Descriptive values for 26 SCN graph-theory features across training, validation, and test sets. Values are: mean (std). Missing data was due to feature generation failure (see Methods): training set 96\% complete; validation 94\%; test 92\%. \newline Notes: Centrality = Betweenness centrality; Clustering = Clustering coefficient.}\label{Table-1}
{\centering
\begin{tabular}{|r|c|c|c|}
 \hline 
\multirow{ 2}{*}{\textbf{Whole Network Features}} & \textbf{Training} & \textbf{Validation} & \textbf{Test} \\
 & (N=3579 of 3739) & (N=390 of 415) & (N=4156 of 4515) \\ \hline 
 Small-world   & 1.68 (0.03) & 1.68 (0.02) & 1.68 (0.02)\\ \hline\xrowht{10pt}
\multirow{2}{*} \textbf{Rich Club} 
-- median & 0.29 (0.01) & 0.29 (0.01) & 0.29 (0.01) \\
-- mad & 0.11 (0.03) & 0.11 (0.01) & 0.11 (0.01) \\ \hline\xrowht{10pt}
\multirow{2}{*} \textbf{Path Length} 
-- median & 2.48 (0.03) & 2.48 (0.01) & 2.48 (0.01) \\
-- std & 1.15 (0.03) & 1.15 (0.02) & 1.16 (0.02) \\ \hline\xrowht{10pt}
\multirow{2}{*} \textbf{Degree} 
-- median & 1050 (45) & 1052 (45) & 1053 (40) \\
-- mad & 295 (19) & 294 (15) & 294 (15) \\ \hline\xrowht{10pt}
\multirow{2}{*} \textbf{Centrality} 
-- median & 6584 (171) & 6590 (150) & 6578 (152) \\
-- mad & 5153 (157) & 5157 (118) & 5158 (117) \\ \hline\xrowht{10pt}
\multirow{2}{*} \textbf{Clustering} 
-- median & 0.53 (0.01) & 0.53 (0.01) & 0.53 (0.01) \\
-- mad & 0.063 (0.005) & 0.063 (0.005) & 0.063 (0.005) \\ \hline
\multicolumn{4}{c}{\textbf{Community 1/2/3 features}} \\ \hline \xrowht{10pt}
\multirow{3}{*} \textbf{Avg. Degree} 
-- 1 & 995 (233) & 1004 (232) & 995 (239) \\\xrowht{5pt}
-- 2 & 996 (240) & 997 (235) & 999 (239) \\\xrowht{5pt}
-- 3 & 1019 (242) & 1004 (243) & 1014 (238) \\ \hline
Avg. degree z-score (all) & 0.2 (0.1) & 0.2 (0.1) & 0.2 (0.1)\\ \hline
Avg. path length (all) & 1.5 (0.1) & 1.5 (0.1) & 1.5 (0.1)\\ \hline\xrowht{10pt}
\multirow{3}{*} \textbf{Centrality} 
-- 1 & 6020 (3750) & 6180 (3800) & 6100 (3780) \\\xrowht{5pt}
-- 2 & 6290 (3790) & 6030 (3740) & 6290 (3770) \\\xrowht{5pt}
-- 3 & 6660 (3830) & 6550 (3760) & 6520 (3810) \\ \hline\xrowht{10pt}
\multirow{3}{*} \textbf{Clustering} 
-- 1 & 0.53 (0.06) & 0.53 (0.06) & 0.52 (0.06) \\\xrowht{5pt}
-- 2 & 0.53 (0.06) & 0.53 (0.06) & 0.53 (0.06) \\\xrowht{5pt}
-- 3 & 0.53 (0.06) & 0.53 (0.06) & 0.52 (0.06) \\ \hline
\end{tabular}}
\end{table}

\section{Results}\label{RESULTS}
We included 26 SCN graph-theory features that represent morphological similarity across the cortex. Table~\ref{Table-1} summarises the features we derived from the T1 images, and the level of completeness in each challenge data set (see Section \ref{predmod}).  For the EBM, only three features passed through our Mann-Whitney U test filter (see Methods): small-worldness, betweenness centrality (median), and degree. Even for these features, there was very little difference between the low- and high-rFIQ groups (see Table~\ref{Table-1}), with Cohen's d effect sizes of $-0.11/0.06$ (small-world), $0.07/-0.10$ (degree), and $-0.09/0.009$ (centrality) in the training/validation sets. In light of the opposing effect direction (signs), the model's poor generalisation performance is unsurprising (see Table~\ref{Table-2}).

For the SVR model, two features were most important: small-worldness (weight $w = 11.42$); and clustering coefficient in community 2 ($w = 6.04$). Among the next most important were average path length and other clustering coefficients.

Table~\ref{Table-2} shows our prediction results for both models: mean-squared errors and Pearson's squared correlation coefficient for training and validation. It is clear that both the approaches did not generalise well under validation. Our submission to the challenge (SVR) was positioned near the middle of the testing leaderboard with $\mathrm{MSE}=93.8335$.

\begin{figure}[h]
\begin{center}
\includegraphics[width=0.95\columnwidth]{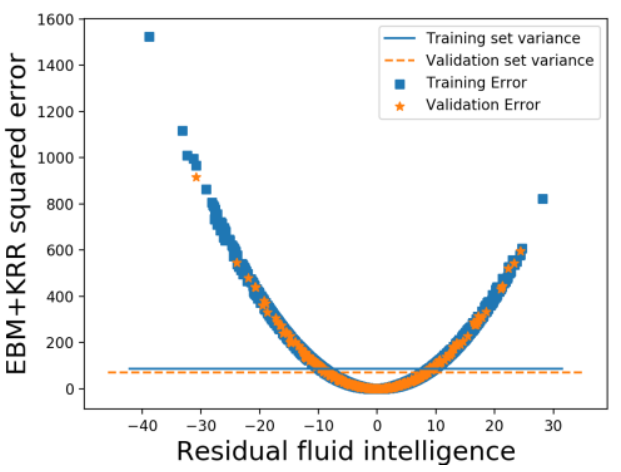}
\caption{Training and validation errors for the EBM approach.}\label{Fig-1}
\end{center}
\end{figure}

\begin{table}[H]
\caption{Mean-squared error (MSE) and correlation for the predictive models. For reference, the variance of the training set was $85.85$ and the validation set was $71.53$.}\label{Table-2}
\begin{center}
\begin{tabular}{c|cc|cc|c}
\hline
\multirow{2}{*}{\textbf{Prediction method}} &  \multicolumn{2}{c|}{Training set}     & \multicolumn{2}{c|}{Validation set}  & Test set \\
                           &  MSE & Correlation &  MSE & Correlation & MSE \\ 
\hline\hline
SVR & 85.82 & 0.02 & 71.19 & 0.01 & \textbf{93.8335} \\ \hline
EBM+KRR & 85.46 & 0.001 & 71.58 & 0.003 & N/A \\ \hline
\end{tabular}
\end{center}
\end{table}

\begin{figure}[H]
\begin{center}
\includegraphics[width=0.9\columnwidth]{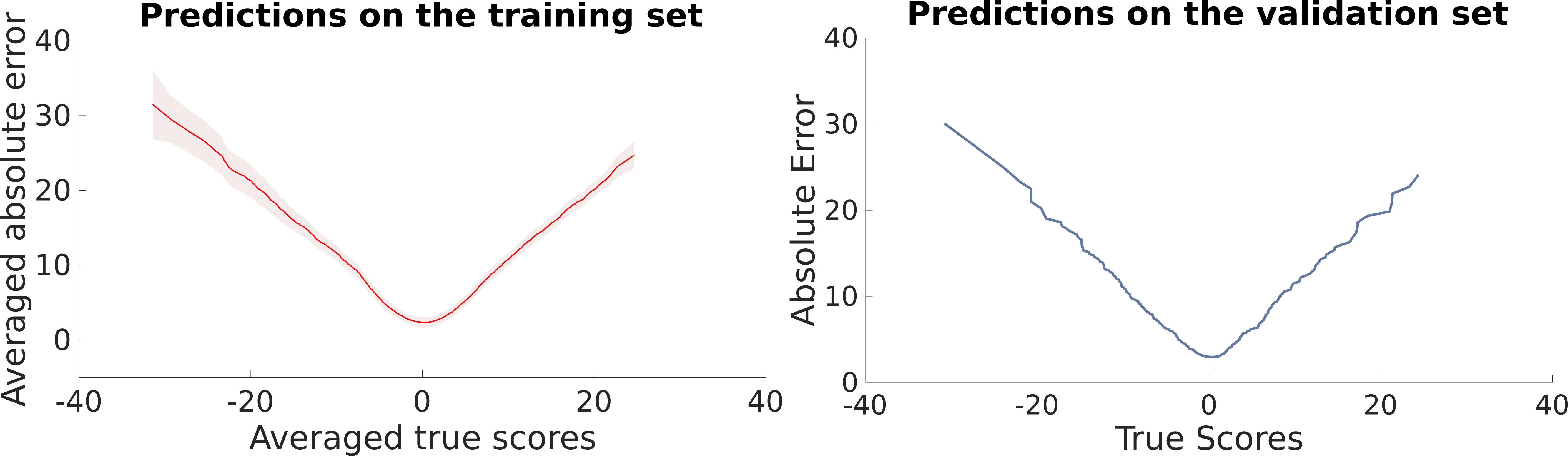}\vspace{5mm}\\
\includegraphics[width=0.9\columnwidth]{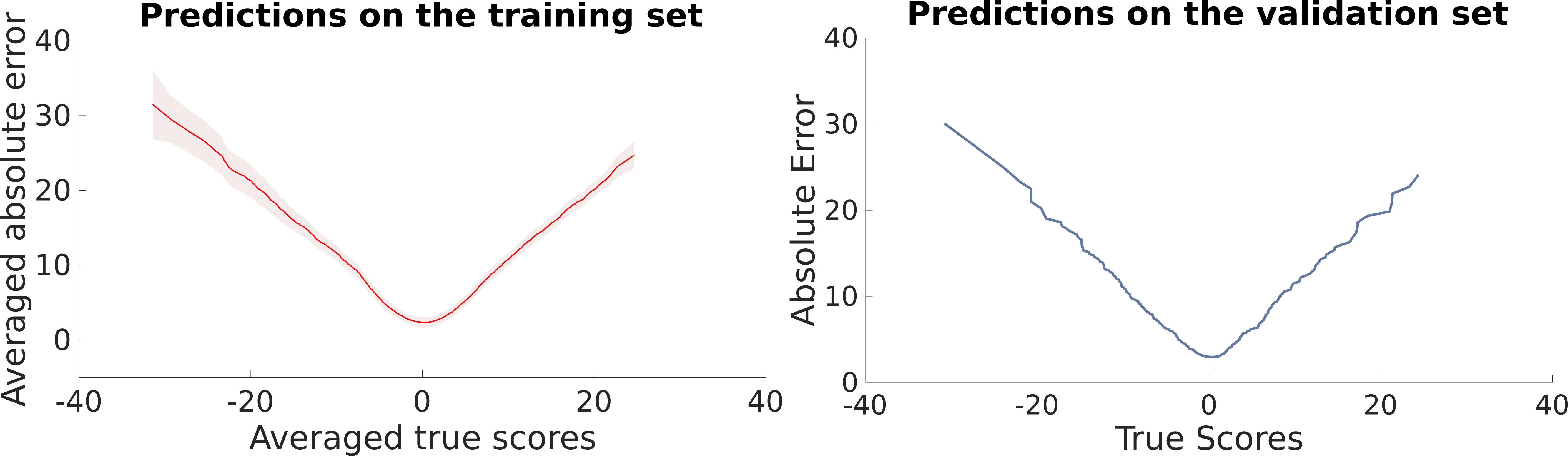}
\caption{SVR prediction errors: (left) training set using 5-fold CV; (right) validation set.}\label{Fig-2}
\end{center}
\end{figure}

\section{Discussion}\label{DISCUSSION}
The ABCD NP Challenge was certainly challenging. Our MSE for predicting residual fluid intelligence was only nominally better than simply predicting zero, i.e., the mean. This implies that the residual fluid intelligence is not explainable by graph theory features derived from structural covariance networks. We found similar results for all combinations of models and features attempted during and after our hackathon --- from basic regression to deep learning. Moreover, the validation leader board (see challenge website) demonstrated that other entries into the challenge had similarly meagre performance improvement on simply predicting the mean. 

While the residualization process precluded the use of models that include covariance of the residualization factors \cite{Schrouff2018} with image-based features, it is difficult to say whether or not this would have improved the results dramatically. Including variables in the residualization procedure that are correlated with the predicted variable is likely to remove important variability in the data leading to predictive models with low performance \cite{Rao2017}.

\section{Conclusion}
Based on our results, and those on the validation leaderboard for the challenge, we are inclined to conclude that structural imaging is probably incapable of predicting more than a couple of points worth of residual fluid intelligence.

\bibliography{ABCD_Refs}

\begin{thebibliography}{10}
\providecommand{\url}[1]{\texttt{#1}}
\providecommand{\urlprefix}{URL }
\providecommand{\doi}[1]{https://doi.org/#1}

\bibitem{Goriounova2019}
Goriounova, N.A., Mansvelder, H.D.: Genes, cells and brain areas of
  intelligence. Frontiers in Human Neuroscience  \textbf{13}, ~44 (2019).
  \doi{10.3389/fnhum.2019.00044}

\bibitem{Foverskov2017}
Foverskov, E., Mortensen, E.L., Holm, A., Pedersen, J.L.M., Osler, M., Lund,
  R.: Socioeconomic position across the life course and cognitive ability later
  in life: The importance of considering early cognitive ability. Journal of
  Aging and Health  (2017). \doi{10.1177/0898264317742810}, pMID: 29254458

\bibitem{Lam2017}
Lam, N.H., Borduqui, T., Hallak, J., Roque, A.C., Anticevic, A., Krystal, J.H.,
  Wang, X.J., Murray, J.D.: {Effects of Altered Excitation-Inhibition Balance
  on Decision Making in a Cortical Circuit Model}. bioRxiv 100347 (2017).
  \doi{10.1101/100347}

\bibitem{Deary2007}
Deary, I.J., Strand, S., Smith, P., Fernandes, C.: {Intelligence and
  educational achievement}. Intelligence  \textbf{35}(1),  13--21 (2007).
  \doi{10.1016/j.intell.2006.02.001}

\bibitem{Deary2013}
Deary, I.J., Pattie, A., Starr, J.M.: {The Stability of Intelligence From Age
  11 to Age 90 Years: The Lothian Birth Cohort of 1921}. Psychological Science
  \textbf{24}(12),  2361--2368 (2013). \doi{10.1177/0956797613486487}

\bibitem{Gottfredson1997}
Gottfredson, L.S.: {Why g matters: The complexity of everyday life}.
  Intelligence  \textbf{24}(1),  79--132 (1997).
  \doi{10.1016/S0160-2896(97)90014-3}

\bibitem{HulshoffPol2006}
{Hulshoff Pol}, H.E., Schnack, H.G., Posthuma, D., Mandl, R.C.W., Baar{\'{e}},
  W.F., {Van Oel}, C., {Van Haren}, N.E., Collins, D.L., Evans, A.C., Amunts,
  K., B{\"{u}}rgel, U., Zilles, K., {De Geus}, E., Boomsma, D.I., Kahn, R.S.,
  Vogt, O.: Genetic contributions to human brain morphology and intelligence.
  Journal of Neuroscience  \textbf{26}(40),  10235--10242 (2006).
  \doi{10.1523/JNEUROSCI.1312-06.2006}

\bibitem{Narr2007}
Narr, K.L., Woods, R.P., Thompson, P.M., Szeszko, P., Robinson, D., Dimtcheva,
  T., Gurbani, M., Toga, A.W., Bilder, R.M.: {Relationships between IQ and
  Regional Cortical Gray Matter Thickness in Healthy Adults}. Cerebral Cortex
  \textbf{17}(9),  2163--2171 (2007). \doi{10.1093/cercor/bhl125}

\bibitem{Choi2008}
Choi, Y.Y., Shamosh, N.A., Cho, S.H., DeYoung, C.G., Lee, M.J., Lee, J.M., Kim,
  S.I., Cho, Z.H., Kim, K., Gray, J.R., Lee, K.H.: {Multiple Bases of Human
  Intelligence Revealed by Cortical Thickness and Neural Activation}. The
  Journal of Neuroscience  \textbf{28}(41),  10323--10329 (2008).
  \doi{10.1523/JNEUROSCI.3259-08.2008}

\bibitem{Karama2011}
Karama, S., Colom, R., Johnson, W., Deary, I.J., Haier, R., Waber, D.P.,
  Lepage, C., Ganjavi, H., Jung, R., Evans, A.C.: {Cortical thickness
  correlates of specific cognitive performance accounted for by the general
  factor of intelligence in healthy children aged 6 to 18}. NeuroImage
  \textbf{55}(4),  1443--1453 (2011). \doi{10.1016/j.neuroimage.2011.01.016}

\bibitem{Jung2007}
Jung, R.E., Haier, R.J.: {The Parieto-Frontal Integration Theory (P-FIT) of
  intelligence: converging neuroimaging evidence.} The Behavioral and Brain
  Sciences  \textbf{30}(2),  135--154 (2007). \doi{10.1017/S0140525X07001185}

\bibitem{Penke2012}
Penke, L., Maniega, S.M., Bastin, M.E., {Vald{\'{e}}s Hern{\'{a}}ndez}, M.C.,
  Murray, C., Royle, N.A., Starr, J.M., Wardlaw, J.M., Deary, I.J.: {Brain
  white matter tract integrity as a neural foundation for general
  intelligence}. Molecular Psychiatry  \textbf{17}, ~1026 (2012).
  \doi{10.1038/mp.2012.66}

\bibitem{Wang2009}
Wang, L., Goldstein, F.C., Veledar, E., Levey, A.I., Lah, J.J., Meltzer, C.C.,
  Holder, C.A., Mao, H.: {Alterations in cortical thickness and white matter
  integrity in mild cognitive impairment measured by whole-brain cortical
  thickness mapping and diffusion tensor imaging}. American Journal of
  Neuroradiology  \textbf{30}(5),  893--899 (2009). \doi{10.3174/ajnr.A1484}

\bibitem{Muetzel2015}
Muetzel, R.L., Mous, S.E., van~der Ende, J., Blanken, L.M., van~der Lugt, A.,
  Jaddoe, V.W., Verhulst, F.C., Tiemeier, H., White, T.: {White matter
  integrity and cognitive performance in school-age children: A
  population-based neuroimaging study}. NeuroImage  \textbf{119},  119--128
  (2015). \doi{10.1016/J.NEUROIMAGE.2015.06.014}

\bibitem{Yu2008}
Yu, C., Li, J., Liu, Y., Qin, W., Li, Y., Shu, N., Jiang, T., Li, K.: {White
  matter tract integrity and intelligence in patients with mental retardation
  and healthy adults}. NeuroImage  \textbf{40}(4),  1533--1541 (2008).
  \doi{10.1016/j.neuroimage.2008.01.063}

\bibitem{Oxtoby2017}
Oxtoby, N.P., Alexander, D.C., {the EuroPOND consortium}: {Imaging plus X:
  multimodal models of neurodegenerative disease.} Current Opinion in Neurology
   \textbf{30}(4),  371--379 (2017). \doi{10.1097/WCO.0000000000000460}

\bibitem{Young2014}
Young, A.L., Alexander, D.C., Oxtoby, N.P., Daga, P., Cash,
  D.M.o.b.o.t.A.D.N.I., Ourselin, S., Schott, J.M., Fox, N.C.: {A data-driven
  model of biomarker changes in sporadic Alzheimer's disease}. Brain
  \textbf{137}(9),  2564--2577 (2014). \doi{10.1093/brain/awu176}

\bibitem{Schrouff2018}
Schrouff, J., Monteiro, J.M., Portugal, L., Rosa, M.J., Phillips, C.,
  Mour{\~{a}}o-Miranda, J.: {Embedding Anatomical or Functional Knowledge in
  Whole-Brain Multiple Kernel Learning Models}. Neuroinformatics
  \textbf{16}(1),  117--143 (2018). \doi{10.1007/s12021-017-9347-8}

\bibitem{Blumberg2018}
Blumberg, S.B., Tanno, R., Kokkinos, I., Alexander, D.C.: Deeper image quality
  transfer: Training low-memory neural networks for 3d images. In: Frangi,
  A.F., Schnabel, J.A., Davatzikos, C., Alberola-L{\'o}pez, C., Fichtinger, G.
  (eds.) Medical Image Computing and Computer Assisted Intervention -- MICCAI
  2018. pp. 118--125. Springer International, Cham (2018)

\bibitem{Bassett2006}
Bassett, D.S., Bullmore, E.: {Small-World Brain Networks}. The Neuroscientist
  \textbf{12}(6),  512--523 (2006). \doi{10.1177/1073858406293182}

\bibitem{Bassett2017}
Bassett, D.S., Sporns, O.: {Network Neuroscience}. Nature Neuroscience
  \textbf{20}(3),  353--364 (2017). \doi{10.1038/nn.4502}

\bibitem{Tijms2013}
Tijms, B.M., M{\"{o}}ller, C., Vrenken, H., Wink, A.M., de~Haan, W., van~der
  Flier, W.M., Stam, C.J., Scheltens, P., Barkhof, F.: {Single-Subject Grey
  Matter Graphs in Alzheimer's Disease}. PLOS ONE  \textbf{8}(3),  e58921
  (2013). \doi{10.1371/journal.pone.0058921}

\bibitem{ABCDWEB}
\url{https://abcdstudy.org/images/Protocol_Imaging_Sequences.pdf}

\bibitem{Casey2018a}
Casey, B.J., et~al.: {The Adolescent Brain Cognitive Development (ABCD) study:
  Imaging acquisition across 21 sites}. Developmental Cognitive Neuroscience
  \textbf{32},  43--54 (2018). \doi{10.1016/j.dcn.2018.03.001}

\bibitem{Akshoomoff2013}
Akshoomoff, N., Beaumont, J.L., Bauer, P.J., Dikmen, S.S., Gershon, R.C.,
  Mungas, D., Slotkin, J., Tulsky, D., Weintraub, S., Zelazo, P.D., Heaton,
  R.K.: {VIII. NIH Toolbox Cognition Battery (CB): composite scores of
  crystallized, fluid, and overall cognition}. Monographs of the Society for
  Research in Child Development  \textbf{78}(4),  119--132 (2013).
  \doi{10.1111/mono.12038}

\bibitem{Hagler2018}
Hagler, D.J., et~al.: {Image processing and analysis methods for the Adolescent
  Brain Cognitive Development Study}. bioRxiv 457739 (2018).
  \doi{10.1101/457739}

\bibitem{Pfefferbaum2018}
Pfefferbaum, A., et~al.: {Altered Brain Developmental Trajectories in
  Adolescents After Initiating Drinking}. American Journal of Psychiatry
  \textbf{175}(4),  370--380 (2018). \doi{10.1176/appi.ajp.2017.17040469}

\bibitem{Rohlfing2010}
Rohlfing, T., Zahr, N.M., Sullivan, E.V., Pfefferbaum, A.: {The SRI24
  multichannel atlas of normal adult human brain structure}. Human Brain
  Mapping  \textbf{31}(5),  798--819 (2010). \doi{10.1002/hbm.20906}

\bibitem{Lawrie2012}
Lawrie, S.M., Tijms, B.M., Willshaw, D.J., Seri{\`{e}}s, P.: {Similarity-Based
  Extraction of Individual Networks from Gray Matter MRI Scans}. Cerebral
  Cortex  \textbf{22}(7),  1530--1541 (2012). \doi{10.1093/cercor/bhr221}

\bibitem{Rubinov2010}
Rubinov, M., Sporns, O.: {Complex network measures of brain connectivity: Uses
  and interpretations}. NeuroImage  \textbf{52}(3),  1059--1069 (2010).
  \doi{10.1016/j.neuroimage.2009.10.003}

\bibitem{Fonteijn2012}
Fonteijn, H.M., Modat, M., Clarkson, M.J., Barnes, J., Lehmann, M., Hobbs,
  N.Z., Scahill, R.I., Tabrizi, S.J., Ourselin, S., Fox, N.C., Alexander, D.C.:
  {An event-based model for disease progression and its application in familial
  Alzheimer's disease and Huntington's disease}. NeuroImage  \textbf{60}(3),
  1880--1889 (2012). \doi{10.1016/j.neuroimage.2012.01.062}

\bibitem{Drucker:1996:SVR:2998981.2999003}
Drucker, H., Burges, C.J.C., Kaufman, L., Smola, A., Vapnik, V.: {Support
  Vector Regression Machines}. In: Proceedings of the 9th International
  Conference on Neural Information Processing Systems. pp. 155--161. NIPS'96,
  MIT Press, Cambridge, MA, USA (1996),
  \url{http://dl.acm.org/citation.cfm?id=2998981.2999003}

\bibitem{Pedregosa2011}
Pedregosa, F., Varoquaux, G., Gramfort, A., Michel, V., Thirion, B., Grisel,
  O., Blondel, M., Prettenhofer, P., Weiss, R., Dubourg, V., Vanderplas, J.,
  Passos, A., Cournapeau, D., Brucher, M., Perrot, M., Duchesnay, {\'{E}}.:
  {Scikit-learn: Machine Learning in Python}. J. Mach. Learn. Res.
  \textbf{12},  2825--2830 (2011)

\bibitem{PRONTO}
\url{http://www.mlnl.cs.ucl.ac.uk/pronto/)}

\bibitem{Schrouff2013}
Schrouff, J., Rosa, M.J., Rondina, J.M., Marquand, A.F., Chu, C., Ashburner,
  J., Phillips, C., Richiardi, J., Mour{\~{a}}o-Miranda, J.: {PRoNTo: pattern
  recognition for neuroimaging toolbox.} Neuroinformatics  \textbf{11}(3),
  319--37 (2013). \doi{10.1007/s12021-013-9178-1}

\bibitem{Rao2017}
Rao, A., Monteiro, J.M., Mourao-Miranda, J., {Alzheimer's Disease Initiative}:
  {Predictive modelling using neuroimaging data in the presence of confounds}.
  NeuroImage  \textbf{150},  23--49 (2017).
  \doi{10.1016/j.neuroimage.2017.01.066}

\end{thebibliography}


\end{document}